\documentstyle[amstex,eqsecnum,aps,multicol,epsfig]{revtex}

\def\be{\begin{eqnarray}}
\def\ee{\end{eqnarray}}

\begin{document}
\title{Generalized measurements via programmable quantum processor}
\author{Mari\'an Ro\v{s}ko$^{1}$, Vladim\'{\i}r Bu\v{z}ek$^{1,2}$,
Paul Robert Chouha$^{3}$, and Mark Hillery$^{3}$}
\address{
${}^{1}$ Research Center for Quantum Information,
Slovak Academy of Sciences, 845 11 Bratislava, Slovakia
\\
${}^{2}$Department of Mathematical Physics, National University of Ireland,
Maynooth, Co. Kildare, Ireland
\\
${}^{3}$Department of Physics and Astronomy, Hunter College of CUNY, 695,
Park
Avenue, New York, NY 10021, U.S.A.
}
\date{September 17, 2003}
\maketitle

\begin{abstract}
We show that it is possible to control the trade-off between
information gain and disturbance in generalized measurements of
qudits by utilizing the  programmable quantum processor. This
universal quantum machine allows us to perform a generalized
measurement on the initial state of the input qudit to construct a
Husimi function of this state. The trade-off between the gain and
the disturbance of the qudit is controlled by the initial state of
ancillary system that acts as a program register for the
quantum-information distributor. The trade-off fidelity does not
depend on the initial state of the qudit.
\end{abstract}


\begin{multicols}{2}

 \section{INTRODUCTION}

Recently  in several  experiments
\cite{Cummins2002,Lamas2002,DeMartini2002} optimal quantum cloning
of qubits \cite{Buzek1996,Gisin1997,Werner1998} has been achieved.
In these experiments the information that was originally encoded
in an (unknown) state $|\Psi\rangle$ of an input qubit has been
distributed between two qubits in a covariant way (i.e., the
fidelity of this information distribution does not depend on the
state of the input qubit). Quantum cloning, viewed as a process of
information distribution, can be considered as one of the basic
tasks of quantum-information processing (QIP). Another important
task of QIP is the application of specific operations (maps) to
the input data. In order to perform either of these tasks, we have
to control the dynamics of the data register. This control can be
achieved by having external forces, which are specified by
classical parameters (e.g. phases and amplitudes of lasers), act
on the quantum system \cite{Harel1999,Lloyd1999,Hladky2000}.
Alternatively, the control of the dynamics of the data register
can be performed on the quantum level, that is the maps induced on
the data register can be completely specified by the quantum state
of a program register in a quantum processor. The action of the
processor is specified by a unitary operator acting on the Hilbert
space of the data and the program register and results in a map
induced on the data
\cite{Nielsen1997,Vidal00,Hillery011,Hillery012,Hillery2002}.

In this paper we will consider a specific model of the quantum
processor - the so-called quantum-information distributor (QID),
which was introduced recently in Ref.\cite{Braunstein2001}. This
covariant quantum processor allows us to distribute quantum
information into several quantum channels as well as to perform
specific quantum operations in each of the channels. This set up
is interesting {\em per se} since it allows us to achieve {\it
quantum} control over quantum systems. In addition, if the
quantum-information distributor is combined with a projective
measurement performed on some of the output channels one can
achieve interesting generalized quantum  (positive operator value
measure) measurements on the input register. In particular, in
this paper we will show how quantum filtering of the original
(input) data register can be realized and how propensities (e.g.,
a Husimi function) of the input register can be easily measured.

Our paper is organized as follows. In Sec.~\ref{sec2} we will
introduce a formal description of a qudit and some basic
operations that can be performed on a single qudit and controlled
rotations that can be performed on two qudits. In Sec.~\ref{sec3}
we will describe the quantum-information distributor and the role
of the programs encoded in states of program qudits.
 Sec.~\ref{sec4} we will be
devoted to a description of generalized measurements and the
reconstruction (measurement) of the Husimi function in a discrete
phase space. In Sec.~\ref{sec5} we will analyze how
positive-operator value measure (POVM) measurement can be realized
with the help of quantum information distributor. We conclude our
paper with some remarks on the noise induced on the input data
qudit due to the projective measurements performed on the program
qudits at the output of quantum-information distributor.

\section{OPERATIONS ON QUDITS}
\label{sec2}
In order to make our discussion self-contained we first present a
brief review of the formalism describing quantum states in a
finite-dimensional Hilbert space. Here we follow the notation introduced
in Ref.~\cite{Buz92}.
Let the $N$-dimensional Hilbert space be spanned by $N$ orthogonal
normalized vectors $|x_{k}\rangle$ or,  equivalently, by $N$ vectors
$|p_{l}\rangle$, $k, l = 0, \ldots, N-1$, where these bases are related
by the discrete Fourier transform
\begin{eqnarray}
\label{2.1}
|x_{k}\rangle &=& \frac{1}{\sqrt{N}} \sum_{l=0}^{N-1} \exp \Bigl(
-i\frac{2\pi}{N}kl \Bigr) |p_{l}\rangle \; ;
\nonumber \\
|p_{l}\rangle &=& \frac{1}{\sqrt{N}} \sum_{k=0}^{N-1} \exp \Bigl(
i\frac{2\pi}{N}kl \Bigr) |x_{k}\rangle \;.
\end{eqnarray}
Without loss of generality, it can be assumed that these bases
consist of sets of eigenvectors of non-commuting operators $\hat
X$ and $\hat P$:
\begin{eqnarray}
\label{2.2}
\hat X |x_{k}\rangle = k|x_{k}\rangle \;, \quad
\hat P |p_{l}\rangle = l|p_{l}\rangle \;,
\end{eqnarray}
that is,
\begin{eqnarray}
\label{2.3}
\hat X = \sum_{k=0}^{N-1} k |x_{k}\rangle \langle x_{k}| \; ;
\qquad
\hat P =  \sum_{l=0}^{N-1} l |p_{l}\rangle \langle p_{l}| \;.
\end{eqnarray}
For instance, we  can assume that the operators $\hat X$ and $\hat P$
are related to a
discrete ``position'' and ``momentum'' of a particle on a ring with a finite
number of equidistant sites \cite{Pegg1988}. Specifically, we can introduce
a length scale, $L$, and two operators, the position
$\hat x$ and the momentum $\hat p$, such that
\begin{eqnarray}
\label{2.4}
\hat x |x_{k}\rangle = x_k|x_{k}\rangle \;, \quad
\hat p |p_{l}\rangle = p_l|p_{l}\rangle \;,
\end{eqnarray}
where
\begin{eqnarray}
\label{2.5}
x_k= L
\sqrt{\frac{2\pi}{N}} k ;
\qquad
p_l=\frac{1}{L} \sqrt{\frac{2\pi}{N}} l\; ,
\end{eqnarray}
where we have used units such that $\hbar=1$. The length, $L$ can,
for example, be taken equal to $\sqrt{1/\omega m}$, where $m$ is
the mass and $\omega$ is the frequency of a quantum
``harmonic''oscillator within a finite dimensional Fock space.

The squared absolute
values of the scalar product of eigenkets (\ref{2.2}) do not depend on the
indices $k$, $l$:
\begin{eqnarray}
\label{2.6}
|\langle x_{k}|p_{l} \rangle |^{2} = 1/N \;,
\end{eqnarray}
which  means that pairs $(k,l)$ form a discrete phase space (i.e.,
pairs $(k,l)$ represent ``points'' of the discrete phase space) on
which (quasi) probability density distributions associated with a
given quantum state can be defined
\cite{Wootters1987,Opatrny1995,Leonhardt1995,Koniorczyk2001,Paz2002}.
Next we introduce operators which shift (cyclicly permute) the
basis vectors \cite{Galetti1988}:
\begin{eqnarray}
\label{2.7}
\hat R_{x}(n)|x_{k}\rangle &=& |x_{(k+n){\rm mod}\, N}\rangle\; ; \nonumber \\
\hat R_{p}(m)|p_{l}\rangle &=& |p_{(l+m){\rm mod}\, N}\rangle\;,
\end{eqnarray}
where the sums of indices are taken modulo $N$ (this summation
rule is considered  throughout this paper, where it is clear we
will not explicitly write the symbol ${\rm mod}\, N$).  For more
about the properties of these operators and the role they play in
the discrete phase space $(k,l)$ see Ref. \cite{Buz95}.

A general single-particle state in the $x$ basis can be expressed
as
\begin{eqnarray}
\label{2.12}
|\Psi\rangle_1 =\sum_{k=0}^{N-1} c_k
|x_k\rangle_1\;; \qquad \sum_{k=0}^{N-1} |c_k|^2 =1\;.
\end{eqnarray}
The basis of maximally entangled two-particle states (the analog
of the Bell basis for spin-$\case{1}{2}$ particles) can be written
as
\begin{eqnarray}
\label{2.13}
|\Xi_{mn}\rangle = \frac{1}{\sqrt{N}} \sum_{k=0}^{N-1} \exp \Bigl(
i\frac{2\pi}{N} mk \Bigr) |x_{k}\rangle|x_{(k-n){\rm mod}\,N}\rangle \,,\!\!\!
\end{eqnarray}
where $m,n=0,\dots,N-1$. We can also rewrite these maximally
entangled states in the $p$ basis:
\begin{eqnarray}
\label{2.14}
|\Xi_{mn}\rangle = \frac{1}{\sqrt{N}} \sum_{l=0}^{N-1} \exp \Bigl(
-i\frac{2\pi}{N} nl \Bigr) |p_{(m-l){\rm mod}\,N}\rangle|p_{l}\rangle\;.\!\!\!
\end{eqnarray}
The states $|\Xi_{mn}\rangle$ form an orthonormal basis
\begin{eqnarray}
\label{2.15}
\langle \Xi_{kl}|\Xi_{mn}\rangle = \delta_{k,m}\delta_{l,n} \;,
\end{eqnarray}
with
\begin{eqnarray}
\label{2.16}
\sum_{m,n=0}^{N-1} |\Xi_{mn}\rangle\langle\Xi_{mn}| = \hat{\openone}\otimes
\hat{\openone}\;.
\end{eqnarray}
In order to prove the above relations we have used the standard relation
$\sum_{n=0}^{N-1}\exp[2\pi i (k-k') n/N] = N \delta_{k,k'}$.

It is interesting to note that the whole set of $N^2$ maximally
entangled states $|\Xi_{mn}\rangle$ can be generated from the
state $|\Xi_{00}\rangle_{23}$ by the action of {\em local} unitary
operations (shifts), e.g.,
\begin{eqnarray}
\label{2.17}
|\Xi_{mn}\rangle_{23} =
\hat{\openone}_2\otimes \hat{R}_x^\dagger(n)
\hat{R}_p(m) |\Xi_{00}\rangle_{23} \;,
\end{eqnarray}
acting just on system $3$ in this particular case.

From the definition of the states $|\Xi_{mn}\rangle_{23}$ it follows that
they are simultaneously eigenstates of the operators $\hat{X}_2 - \hat{X}_3$
and $\hat{P}_2 + \hat{P}_3$:
\begin{eqnarray}
\label{2.18}
(\hat{X}_2 - \hat{X}_3)|\Xi_{mn}\rangle_{23} &=& n |\Xi_{mn}\rangle_{23}\; ;
\nonumber
\\
(\hat{P}_2 + \hat{P}_3)|\Xi_{mn}\rangle_{23} &=& m |\Xi_{mn}\rangle_{23} \;.
\end{eqnarray}
We easily see that for $N=2$ the above formalism reduces to the well-known
spin-$\case{1}{2}$ particle (qubit) case.

Now we introduce generalizations of the two-qubit controlled-NOT
(CNOT) gate (see also Ref.~\onlinecite{alber}).  In the case of
qubits the CNOT gate is represented by a two-particle operator
such that if the first (control) particle labeled $a$ is in the
state $|0\rangle$ nothing ``happens'' to the state of the second
(target) particle labeled $b$. If, however, the control particle
is in the state $|1\rangle$ then the state of the target is
``flipped'', i.e., the state $|0\rangle$ is changed into the state
$|1\rangle$ and vice versa. Formally we can express the action of
this CNOT gate  as a two-qubit operator of the form
\begin{eqnarray}
\label{2.19}
\hat{D}_{ab}=\sum_{k,m=0}^1 |k\rangle_a\langle k|\otimes
|(m+k){\rm mod}\,2\rangle_b\langle m|\;.
\end{eqnarray}
We note that in principle one can introduce an operator
$\hat{D}^\dagger_{ab}$ defined as
\begin{eqnarray}
\label{2.20}
\hat{D}^\dagger_{ab}=\sum_{k,m=0}^1 |k\rangle_a\langle k|\otimes
|(m-k){\rm mod}\,2\rangle_b\langle m|\;.
\end{eqnarray}
In the case of qubits these two operators are equal. This is not
the case when the dimension of the Hilbert space is larger than
two \cite{alber}. Let us generalize the above definition of the
operator $\hat{D}$ for $N>2$. Before doing so, we shall simplify
our notation. Because we will work mostly in the $x$ basis we
shall use the notation $|x_k\rangle\equiv|k\rangle$, where it may
be done so unambiguously. With this in mind we now write
\begin{eqnarray}
\label{2.21}
\hat{D}_{ab}=
\sum_{k,m=0}^{N-1} |k\rangle_a\langle k|\otimes
|(m+k){\rm mod}\,N\rangle_b\langle m|\;.
\end{eqnarray}
From  definition (\ref{2.21}) it follows that the operator
$\hat{D}_{ab}$ acts on the basis vectors as \be
\hat{D}_{ab}|k\rangle|m\rangle = |k\rangle|(k+m) {\rm mod}\,
N\rangle\;, \label{2.22} \ee which means that this operator is
equal to the conditional adder \cite{vedral96,pittenger} and can
be performed with the help of a simple quantum network as
discussed in Ref. \cite{vedral96}.

\begin{figure}[tbp]
\centerline {\epsfig{width=8.0cm,file=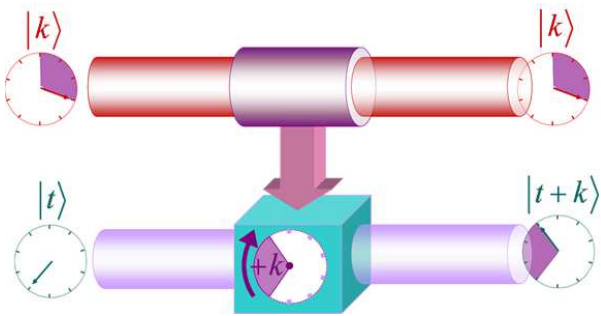}}
\medskip
\caption{Schematic description of the two-qudit conditional-shift
gate.
}
\label{fig1}
\end{figure}

If we take into account the definition of the shift operator
$\hat{R}_x(n)$ given by Eq.~(\ref{2.7})
and the definition of the position and momentum operators $\hat{x}$
and $\hat{p}$ given by Eq.~(\ref{2.4})
we can rewrite the operator
$\hat D_{ab} $ as:
\begin{eqnarray}
\label{2.23}
\hat{D}_{ab}&=&\sum_{k,m=0}^{N-1} |k\rangle_a\langle k|\otimes
\hat{R}_x^{(b)}(k)|m\rangle_b\langle m|
\\
\nonumber
&\equiv&
\sum_{k=0}^{N-1} |k\rangle_a\langle k|\otimes
\hat{R}_x^{(b)}(k) \;,
\end{eqnarray}
and analogously
\begin{eqnarray}
\label{2.24}
\hat{D}_{ab}^\dagger&=&
\sum_{k,m=0}^{N-1} |k\rangle_a\langle k|\otimes
|(m-k){\rm mod}\,N\rangle_b\langle m|
\\
\nonumber
&\equiv&
\sum_{k=0}^{N-1} |k\rangle_a\langle k|\otimes
\hat{R}_x^{(b)}(-k) \;,
\end{eqnarray}
where the subscripts $a$ and $b$ indicate on which Hilbert space
the given operator acts. Now we see that for $N>2$ the two
operators $\hat D$ and $\hat{D}^\dagger$ do differ; they describe
conditional shifts in opposite directions. We see that the
generalization of the CNOT operator are the {\em conditional
shifts}. The amount by which the target (in our case particle $b$)
is shifted depends on the state of the control particle ($a$) [for
a pictorial representation of this gate see Fig.~\ref{fig1}]

\section{Quantum Information Distributor}
\label{sec3} As shown in Ref.~\cite{Braunstein2001} quantum
control over the quantum information can be achieved with  the
help of a quantum ``machine'', the so-called quantum information
distributor (QID). The machine takes as an input a system qudit
prepared in an unknown state $|\Psi\rangle_1$ and two ancilla
qudits prepared in the state $|\Theta\rangle_{23}$ that play the
role of quantum program (i.e., the CP map that has to be performed
on the system qubit is encoded in this state). The action of the
QID itself is described by a unitary operator $U_{123}$ acting on
the Hilbert space that is a tensor product of the three qudits
under consideration. This unitary operator can be expressed as a
sequence of four controlled shifts $D_{kl}$, i.e., \be
 \hat{U}_{123}=\hat{D}_{31}\hat{D}_{21}^{\dagger}\hat{D}_{13}\hat{D}_{12}\, .
\label{3.1}
\ee
The flow of information in the quantum distributor,
as described by the unitary operator (\ref{3.1}), is governed by the
preparation of the distributor itself, i.e., by the choice of the
program state $|\Theta\rangle_{23}$. In other words,
we imagine the transformation
(\ref{3.1}) as a universal  ``processor'' or distributor and the
state $|\Theta\rangle_{23}$ as ``program''
through which the information flow is controlled.

\begin{figure}[tbp]
\centerline {\epsfig{width=8.0cm,file=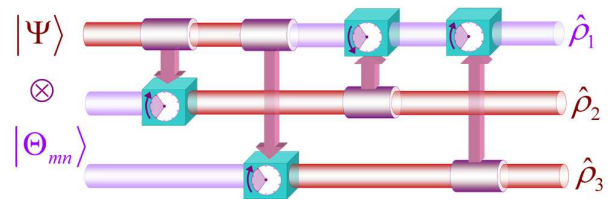}}
\medskip
\caption{Logical network for the quantum-information distributor.
The network is composed of four conditional-shift gates. }
\label{fig2}
\end{figure}

We present the logical network for the QID in Fig.~\ref{fig2} The
output state of the three-particle system after the four
controlled shifts are applied is
\begin{eqnarray}
\label{3.3}
|\Omega^{(out)}\rangle_{123}=
\hat{D}_{31}\hat{D}_{21}^\dagger\hat{D}_{13}\hat{D}_{12}
|\Psi\rangle_1|\Theta\rangle_{23}\;.
\end{eqnarray}
Note that the QID is covariant
with respect to any choice of the state $|\Psi\rangle_{1}$
of data register (for more details see Ref.~\cite{Braunstein2001}).

\subsection{Factorized program states}
\label{subsec3a} Let us first assume that the two program qudits
are in a pure state \be
|\Theta\rangle_{23}=|x_m\rangle_2|p_n\rangle_3\; . \label{3.4}
\ee After the action of the QID the state
$|\Omega\rangle_{123}=|\Psi\rangle_1\otimes |\Theta\rangle_{23}$
transforms as
 \be
 |\Omega^{(out)}\rangle_{123} &=& U_{123}|\Psi\rangle_1
|x_m\rangle_2|p_n\rangle_3 \nonumber
\\
&=&
 \left[\hat{R}_x(m)\hat{R}_p^{\dagger}(n)|\Psi \rangle_2\right]
  \otimes |\Xi_{nm} \rangle_{31}\; .
\label{3.5}
 \ee
So we can observe two actions of the QID on the input state:
First, the state of the original qudit has been totally copied on
the state of the second qudit. Simultaneously, the second qudit
undergoes two rotations described by the operator
$\hat{R}_x(m)\hat{R}_p^{\dagger}(n)$, where the values of the
rotations are uniquely determined by the program state. Finally,
the two remaining qudits (labeled as $1$ and $3$) became maximally
entangled as the result of the action of the QID.

\subsection{Maximally entangled program states}
\label{subsec3b}
Let us  assume that the QID state $|\Theta\rangle_{23}$ is initially
prepared in the
maximally entangled state $|\Xi_{mn}\rangle_{23}$ given by Eq.~(\ref{2.13})
Taking the original system to be prepared in the state
$|\Psi\rangle_1$, i.e. the three qudits at the input are in the state
\be
|\Omega\rangle_{123}=|\Psi\rangle_1\otimes|\Xi_{mn}\rangle_{23}
\label{3.6}
\ee
we find after the QID transformation the expression for the state
vector of the three qudits
\be
|\Omega^{(out)}\rangle_{123} = \left[\hat{R}^{\dagger}_{x}(n)
 \hat{R}^{\dagger}_{p}(m)|\Psi\rangle_1 \right]
\otimes|\Xi_{mn}\rangle_{23} \; .
\label{3b.2}
\ee
We see that if the program register is initially prepared
in the maximally entangled
state then the information encoded in the input state of the first (system)
qudit will remain in this qudit, but the QID will induce a specific rotation
on this qudit that is uniquely determined by the maximally entangled state
of the program qudits. Interestingly enough, the program state is not
changed at all in this case.

\subsection{Superposition of  program states}
\label{subsec3c} The complete set of maximally entangled states
Eq. (\ref{2.13}) is a basis for the two-qudit Hilbert space.
Therefore, an arbitrary program state can be written in the form
\be |\Theta\rangle_{23}=\sum_{m,n=0}^{N-1}d_{mn}|\Xi_{mn}\rangle\;
. \label{3.8} \ee and the corresponding evolution of the QID
results in the state \be \label{3.9}
 \hat{U}|\Omega\rangle &=& \sum_{m,n=0}^{N-1}d_{mn}\hat{R}^{\dagger}_{x}(n)
 \hat{R}^{\dagger}_{p}(m)|\Psi\rangle_1 \otimes|\Xi_{mn}\rangle_{23} \\
\label{3.10}
 &=&\sum_{m,n=0}^{N-1}\tilde{d}_{mn}
 \left[\hat{R}^{\dagger}_{p}(n)\hat{R}_{x}(m)
|\Psi\rangle_2\right] \otimes|\Xi_{nm}\rangle_{13}\; , \ee where
$\tilde{d}_{mn}$ is the Fourier transformation of the coefficients
$d_{mn}$: \be \label{3.11} \tilde{d}_{mn} = \frac{1}{N}
\sum_{k,l=0}^{N-1} d_{kl} ~\exp\left[i \frac{2 \pi}{N}
(km+ln)\right] = \textit{F}(d_{mn})\; . \ee This last result is
not surprising, since the complete set of factorized states
$|x_m\rangle|p_n\rangle$ forms another orthonormal basis for the
program space. If the program space is expanded in this basis Eq.\
(\ref{3.10}) immediately results. What is interesting is that the
program state that induces a specific operation on the first qudit
performs an analogous (though not identical) operation on the
second qudit. To see this we present the reduced density operators
of these two qudits at the output of the QID [see
Eqs.~(\ref{3.9}) and (\ref{3.10})]: \be \label{3.12} \hat{\rho}_1
&=& \sum_{m,n=0}^{N-1} |d_{mn}|^2 \hat{R}^{\dagger}_x(n)
\hat{R}^{\dagger}_p(m)|\Psi\rangle\langle\Psi|\hat{R}_p(m)\hat{R}_x(n)
\; ;
\\
\hat{\rho}_2 &=& \sum_{m,n=0}^{N-1} |\tilde{d}_{mn}|^2
\hat{R}_x(m)
\hat{R}^{\dagger}_p(n)|\Psi\rangle\langle\Psi|\hat{R}_p(n)\hat{R}_x^{\dagger}
(m) \; . \label{3.13} \ee We will use this property of the QID and
it application as a measurement device that realizes a generalized
measurement in our further analysis of the QID.

\subsection{The case of qubits}
We have seen that for qudits, there are two special bases that cause a
set of operations to be performed on the input data state so that at the
output, the transformed data state is disentangled from the output
program state.  In the case of qubits there is a third.

In this section let us change our notation to connect it to that
usually employed for two-state systems.  We shall denote the
states $|x_{0}\rangle$ and $|x_{1}\rangle$ by $|0\rangle$ and
$|1\rangle$, respectively.  The states $|p_{0}\rangle$ and
$|p_{1}\rangle$ can then be expressed as
\begin{eqnarray}
|p_{0}\rangle & = & \frac{1}{\sqrt{2}}(|0\rangle +|1\rangle ) \, ;\nonumber
\\
|p_{1}\rangle & = & \frac{1}{\sqrt{2}}(|0\rangle -|1\rangle ) .
\end{eqnarray}
The maximally entangled states $|\Xi_{mn}\rangle$ are just the Bell states.
The actions of the product and maximally entangled program states can
be expressed as
\begin{eqnarray}
U_{123}|\Psi\rangle_{1}|\Xi_{00}\rangle_{23} & = & |\Psi\rangle_{1}
|\Xi_{00}\rangle_{23}\, ; \nonumber \\
U_{123}|\Psi\rangle_{1}|\Xi_{10}\rangle_{23} & = &( \sigma_{z}|\Psi\rangle_{1})
|\Xi_{10}\rangle_{23}\, ; \nonumber \\
U_{123}|\Psi\rangle_{1}|\Xi_{01}\rangle_{23} & = & (\sigma_{x}|\Psi\rangle_{1})
|\Xi_{01}\rangle_{23}\, ; \nonumber \\
U_{123}|\Psi\rangle_{1}|\Xi_{11}\rangle_{23} & = & -i(\sigma_{y}
|\Psi\rangle_{1})|\Xi_{11}\rangle_{23} ,
\end{eqnarray}
and
\begin{eqnarray}
U_{123}|\Psi\rangle_{1}|0\rangle_{2}|p_{0}\rangle_{3} & = &
|\Psi\rangle_{2}|\Xi_{00}\rangle_{13}\, ; \nonumber \\
U_{123}|\Psi\rangle_{1}|0\rangle_{2}|p_{1}\rangle_{3} & = & (\sigma_{z}
|\Psi\rangle_{2}) |\Xi_{10}\rangle_{13}\, ; \nonumber \\
U_{123}|\Psi\rangle_{1}|1\rangle_{2}|p_{0}\rangle_{3} & = & (\sigma_{x}
|\Psi\rangle_{2})|\Xi_{01}\rangle_{13}\, ; \nonumber \\
U_{123}|\Psi\rangle_{1}|1\rangle_{2}|p_{1}\rangle_{3} & = & i(\sigma_{y}
|\Psi\rangle_{2})|\Xi_{11}\rangle_{13} .
\end{eqnarray}
There is now a second product basis that causes the transformed data
state to emerge from output $3$.  We have that
\begin{eqnarray}
U_{123}|\Psi\rangle_{1}|p_{0}\rangle_{2}|0\rangle_{3} & = &
|\Psi\rangle_{3}|\Xi_{00}\rangle_{12}\, ; \nonumber \\
U_{123}|\Psi\rangle_{1}|p_{1}\rangle_{2} |0\rangle_{3} & = & (\sigma_{z}
|\Psi\rangle_{3}) |\Xi_{10}\rangle_{12}\, ; \nonumber \\
U_{123}|\Psi\rangle_{1}|p_{0}\rangle_{2} |1\rangle_{3} & = & (\sigma_{x}
|\Psi\rangle_{3})|\Xi_{01}\rangle_{12}\, ; \nonumber \\
U_{123}|\Psi\rangle_{1}|p_{1}\rangle_{2} |1\rangle_{3} & = & i(\sigma_{y}
|\Psi\rangle_{3})|\Xi_{11}\rangle_{12} .
\end{eqnarray}

The additional basis suggests that it would be useful to examine
program states that are superpositions of three states, one from
each basis.  Perhaps the simplest of these is the one that is a
superposition of the states corresponding to the identity operator
\begin{equation}
|\Theta\rangle_{23}=\alpha |\Xi_{00}\rangle_{23} + \beta
|0\rangle_{2} |p_{0}\rangle_{3}+\gamma
|p_{0}\rangle_{2}|0\rangle_{3} ,
\end{equation}
where $|\Xi_{00}\rangle_{ab}=(|00\rangle+|11\rangle)/\sqrt{2}$.
 The normalization condition for this state is
\begin{equation}
|\alpha +\beta |^{2}+|\alpha +\gamma |^{2}+|\beta +\gamma |^{2}=2 .
\end{equation}
We hope that this state will lead to an output that consists of
three approximate copies of the input data state,
$|\Psi\rangle_{1}$, however, we find that this is not what
happens. The reduced density matrix of the first output is
\begin{eqnarray}
\hat{\rho}_{1}&=&\left[\left\vert\alpha +\frac{(\beta +\gamma
)}{2}\right\vert^{2}-\frac{|\beta +\gamma
|^{2}}{4}\right]\hat{\rho} +\frac{|\beta +\gamma
|^{2}}{2}\hat{\openone}
\nonumber \\
&-&\frac{(\beta\gamma^{\ast}+\beta^{\ast}\gamma
)}{2}\sigma_{y}\hat{\rho}\sigma_{y} ,
\end{eqnarray}
where we used the notation $\hat{\rho}=|\Psi\rangle\langle\Psi|$.
Similar results are obtained for the reduced density matrices
of outputs $2$ and $3$. As can be seen, while the first two terms
are, in fact, an approximate copy of the input state, this is
disturbed by the last term.  Note that if either $\beta$ or
$\gamma$ is zero, this term disappears and the device behaves as
an approximate cloner.

A somewhat more successful example is given by a program state consisting
of states each of which corresponds to a different operation, e.g.
\begin{equation}
|\Theta\rangle_{23}=\alpha |\Xi_{00}\rangle_{23} + \beta
|0\rangle_{2} |p_{1}\rangle_{3}+\gamma
|p_{0}\rangle_{2}|1\rangle_{3} ,
\end{equation}
with the normalization condition
\begin{equation}
|\alpha +\beta |^{2}+|\alpha +\gamma |^{2}+|\beta -\gamma |^{2} =2 .
\end{equation}
The first state in the superposition produces $|\Psi\rangle$ in
output $1$, the second $\sigma_{z}|\Psi\rangle$ in output $2$, and
the third $\sigma_{x}|\Psi\rangle$ in output $3$.  The
single-qubit reduced density matrices resulting from this program
state are
\begin{eqnarray}
\hat{\rho}_{1} & = & \left[\left\vert\alpha +
\frac{(\beta+\gamma)}{2} \right\vert^{2} -
 \frac{|\beta -\gamma|^{2}}{4}\right]\hat{\rho} +\frac{|\beta
-\gamma |^{2}}{2}\hat{\openone} \, ; \\
\hat{\rho}_{2} & = & \left[\left\vert\beta +\frac{(\alpha -\gamma
)}{2}\right\vert^{2}-\frac{|\alpha +\gamma
|^{2}}{4}\right]\sigma_{z}\hat{\rho} \sigma_{z}
+\frac{|\alpha +\gamma |^{2}}{2}\hat{\openone} \, ;\nonumber \\
\hat{\rho}_{3} & = & \left[\left\vert\gamma +\frac{(\alpha -\beta
)}{2}\right\vert^{2}-\frac{ |\alpha +\beta
|^{2}}{4}\right]\sigma_{x}\hat{\rho}\sigma_{x} +\frac{|\alpha
+\beta |^{2}}{2}\hat{\openone} \nonumber \, .
\end{eqnarray}
In this case we do obtain an approximate version of $|\Psi\rangle$
in output $1$, an approximate version of $\sigma_{z}|\Psi\rangle$
in output $2$, and an approximate version of
$\sigma_{x}|\Psi\rangle$ in output $3$.  A simple way to see how
the accuracy of the approximations in the different outputs is
constrained, is to define the fidelities
\begin{eqnarray}
F_{1} & = & \langle\Psi |\hat{\rho}_{1}|\Psi\rangle\, ; \nonumber \\
F_{2} & = & \langle\sigma_{z}\Psi
|\hat{\rho}_{2}|\sigma_{z}\Psi\rangle\, ;
\nonumber \\
F_{3} & = & \langle\sigma_{x}\Psi
|\hat{\rho}_{3}|\sigma_{x}\Psi\rangle ,
\end{eqnarray}
and to note that $F_{1}+F_{2}+F_{3}=2$.  Each of the fidelities
lies between $1/2$ and $1$. Noting that a completely noisy output
of $ \hat{\openone}/2$, containing no information about the input,
corresponds to a fidelity of $1/2$, we see that if one of the
fidelities is $1$, containing perfect information about the input,
the others are  $1/2$, and contain no information.  Thus, we have
a kind of conservation of information, the more accurate one
output becomes, the less accurate the others become in order to
compensate.  If the fidelities are equal, then each is equal to
$2/3$. This is the fidelity of state estimation, and hence
cloning, that would be achieved by simply measuring the input
qubit.

\section{Quantum propensities}
\label{sec4}
According to W\'{o}dkiewicz \cite{Wod}, propensity means the tendency
(or probability) of a measured
object to take up certain states prescribed by a measuring device. Let
the  measuring device - the so called quantum ruler -
be  in a pure state $|\Phi\rangle$. The quantum-ruler state can be ``shifted'' by an action of some
generalized displacement operator $\hat D(g)$, where $g$ is an
element of a group $G$. If the measured system is in a pure state
$|\Psi \rangle$, then its probability
to be in the ruler state shifted by $g$ (i.e., the propensity)
is
\begin{eqnarray}
\label{4.1}
P_{\Phi ,\Psi}(g) &=& |\langle \Psi |\hat D(g)| \Phi \rangle |^{2} ,
\end{eqnarray}
whereas if the system is in a mixed state described by the density operator
$\hat \rho$, the propensity is
\begin{eqnarray}
\label{4.2}
P_{\Phi, \rho}(g) &=& \mbox{Tr} \left( \hat \rho \hat D(g) |\Phi \rangle
\langle \Phi | \hat D^{+}(g)  \right) .
\end{eqnarray}
In our case, that of a finite dimensional Hilbert space, the group
$G$ will be formed by discrete translations on a torus: if $g_{1}
\equiv (n_{1}, m_{1})$ and $g_{2} \equiv (n_{2}, m_{2})$ are
elements of $G$, then their group product is $g_{1}g_{2} \equiv
\left( (n_{1}+n_{2}) \mbox{mod} N, (m_{1}+m_{2}) \mbox{mod} N
\right)$. The corresponding displacement operator is then given by
the expression $\hat R_{x}(n) \hat R_{p}(m)$.
 We see that while the displacement
is not a representation of the group $G$ in the Hilbert space
under consideration, nevertheless it is representation of this
group in a ray space, which enables us to define the propensity
uniquely. For a pure state $|\Psi \rangle$ we can write the
propensity in the form (see Ref. \cite{Buz95}):
\begin{eqnarray}
\label{4.3}
P_{\Phi ,\Psi }(n, m) &=& |\langle \Psi |\hat R_{x}(n) \hat R_{p}(m)
|\Phi \rangle |^{2}.
\end{eqnarray}
In the case of a statistical mixture described by the density operator
 $\hat \rho$ the corresponding propensity reads
\begin{eqnarray}
\label{4.4}
P_{\Phi ,\rho}(n,m) &=& \mbox{Tr} \left( \hat \rho \hat R_{x}(n) \hat
R_{p}(m) |\Phi \rangle \langle \Phi |
\hat R_{p}^\dagger(n) \hat R_{x}^\dagger(m)
\right) .
\end{eqnarray}

\subsection{Propensities and POVM measurements}
The propensities as defined above are in fact results of so-called
generalized (positive operator value measure - POVM) measurements
(e.g., see Ref.~\cite{Nielsen}). To see this let us recall that
 \be
\label{4.5}
 \hat{F}_{mn}= \hat{R}_x(n)\hat{R}_p(m)
 |\Phi\rangle\langle\Phi|\hat{R}_p^{\dagger}(m)\hat{R}_x^{\dagger}(n)
 \ee
where $|\Phi\rangle$ is a ruler state are {\em positive} operators
and they fulfill the condition
 \be
\label{4.6}
 \sum_{mn} \hat{F}_{mn} = N \hat{\openone}
 \ee
 So the operators $\hat{F}_{mn}$ (or more specifically the operators
$\hat{f}_{mn}=\hat{F}_{mn}/N$) form a complete set  that can be
used for a complete measurement of the state of a qudit. We note
that other operators of the form (\ref{4.5}), e.g.,
 \be
\label{4.7}
 \hat{F}_{mn}= \hat{R}_x(m)\hat{R}_p^{\dagger}(n)
 \hat{\rho}\hat{R}_p(n)\hat{R}_x^{\dagger}(m)
 \ee
also realize a POVM measurement.

\subsection{$Q$ function in discrete phase space}
\label{sec4a} In an analogy with a continuous $(q,p)$ phase space,
where the $Q$ function (Husimi function) is defined as the
propensity of a state to be in the
 vacuum  state, we define the discrete
$Q$ function as propensity (\ref{4.1})
\begin{eqnarray}
\label{4.8}
Q(n,m) &\equiv & P_{\Phi ,\rho}(n,m),
\end{eqnarray}
with the quantum ruler being in a ``vacuum'' state. The problem is how
to define a vacuum state corresponding to a finite-dimensional
Hilbert space.

Before specifying the ruler state, we  will mention several
properties of  discrete $Q$ functions. If we assume that the ruler
state $|\Phi \rangle$ is chosen (i.e., the vacuum state is
specified) then the $Q$ function  has the following
properties:\newline {\bf (i)} it is uniquely defined;\newline {\bf
(ii)} it is non-negative;\newline {\bf (iii)} it is normalized to
$N$
\begin{eqnarray}
\label{4.9}
\sum_{n,m} Q(n,m) &=& N;
\end{eqnarray}
{\bf (iv)} for {\em properly} chosen ruler states $|\Phi \rangle$
the information about a  system state can be completely
reconstructed from the corresponding $Q$ function.

\subsection{Ruler state}
In analogy with the continuous limit, where the ruler state associated
with a Husimi function is the ground (vacuum) state of the harmonic
oscillator, let us consider following requirements
on the ruler state: {\bf (i)} it
should be in some sense centered at
origin of phase space [i.e., the point $(0,0)$], {\bf (ii)}
it should be ``symmetric'' with regards to the quantities $X$ and $P$,
i.e., its wave function should have similar form in both
representations (perhaps up to scalings),
and {\bf (iii)} it should be in some sense a minimum uncertainty
state, which means  that
 in the phase space it should be represented by a peak which is
as narrow as possible. As shown in Ref.~\cite{Opatrny1995} all the
above properties are fulfilled by the ground state of the
Hamiltonian
 \be
\label{4.10}
  \hat{H}_0 = -\cos(\frac{2\pi}{N}\hat{X}) -\cos(\frac{2\pi}{N}\hat{P})\; .
 \ee
We will use this ground state as the ruler state in our forthcoming
considerations.

\section{POVM MEASUREMENT VIA QID}
\label{sec5}
Let us now study the action of the quantum information distributor
when the two ancillary qudits are prepared in a superposition state
 \be
\label{5.1}
 |\Theta \rangle_{23} =\left(\alpha|\Xi_{00}\rangle_{23}+
 \beta|x_m \rangle_{2} |p_n \rangle_{3}\right)\;
 ,
 \ee
with the two real amplitudes $\alpha$ and $\beta$ satisfying the
normalization condition
 \be
  \alpha^2+\beta^2+\frac{2\alpha\beta}{N}
\cos\left(\frac{2\pi}{N}nm\right)=1\; .
\label{5.2}
 \ee
With this program state the QID acts on the input data qudit
$|\Psi\rangle_1=\sum_k c_k |x_k\rangle$
so that at the output the three qudits are in the
following states:
 \be
\label{5.3}
 \hat{\rho}_1&=&(1-\beta^2)\hat{\rho}+\frac{\beta^2}{N}\hat{\openone}\; ; \\
 \hat{\rho}_2&=&(1-\alpha^2)\hat{R}_x(m)\hat{R}_p^{\dagger}(n)\hat{\rho}
\hat{R}_p(n)\hat{R}_x^{\dagger}(m)+ \frac{\alpha^2}{N}\hat{\openone}\; ;
\label{5.4}
\\
 \hat{\rho}_3&=&(1-\alpha^2-\beta^2) \hat{R}_x(m)\hat{R}_p(n) \hat{\rho}^{\rm T}
 \hat{R}_p^{\dagger}(n)\hat{R}_x^{\dagger}(m) \nonumber\\
 &&+\frac{\alpha^2+\beta^2}{N}\hat{\openone} \; ,
\label{5.5}
 \ee
where $\hat{\rho}=|\Psi\rangle\langle\Psi|$ and $\hat{\rho}^{\rm
T}$ is the transpose of the density operator
$\hat{\rho}=\sum_{k,k^{\prime}} c_k
c^*_{k^\prime}|x_k\rangle\langle x_{k^\prime}|$. That is, in the
basis $|x_k\rangle$ the transposed density operator reads
$\hat{\rho}^{\rm T}=\sum_{k,k^{\prime}} c^*_k
c_{k^\prime}|x_k\rangle\langle x_{k^\prime}|$.

The action of the QID discussed earlier,
allows us to  reconstruct partially the state
 of the measured system without a  total ``destruction''
of  the state of the data register. Specifically, from Eq.~(5.3) it follows
that the entangled component of the program register (represented by the
state $|\Xi_{00}\rangle_{23}$) dictates how ``much''
of the original information
encoded in the qudit 1 is transferred from the data register to the program
register at the output of the QID. For instance, if the amplitude $\alpha$
is equal to unity (i.e., $\beta=0$) then the data register is not perturbed
at all, and no information is transferred.
On the other hand, for $\alpha <1$ some of the information
from the data is transferred to the program at the expense of noise
introduced into the data register. The trade-off between the information
transfer and the noise introduced into the data register is nicely seen from
Eq.~(5.3). The amount of noise that is transferred into the first (data)
qudit is dictated by the
amplitude $\beta$ that weights the factorizable contribution to the program
state, i.e. $|x_n\rangle_2|p_m\rangle_3$. Moreover this specific state also
determines  operations (rotations)  that are performed on program qudits.

\begin{figure}[tbp]
\centerline {\epsfig{width=10.0cm,file=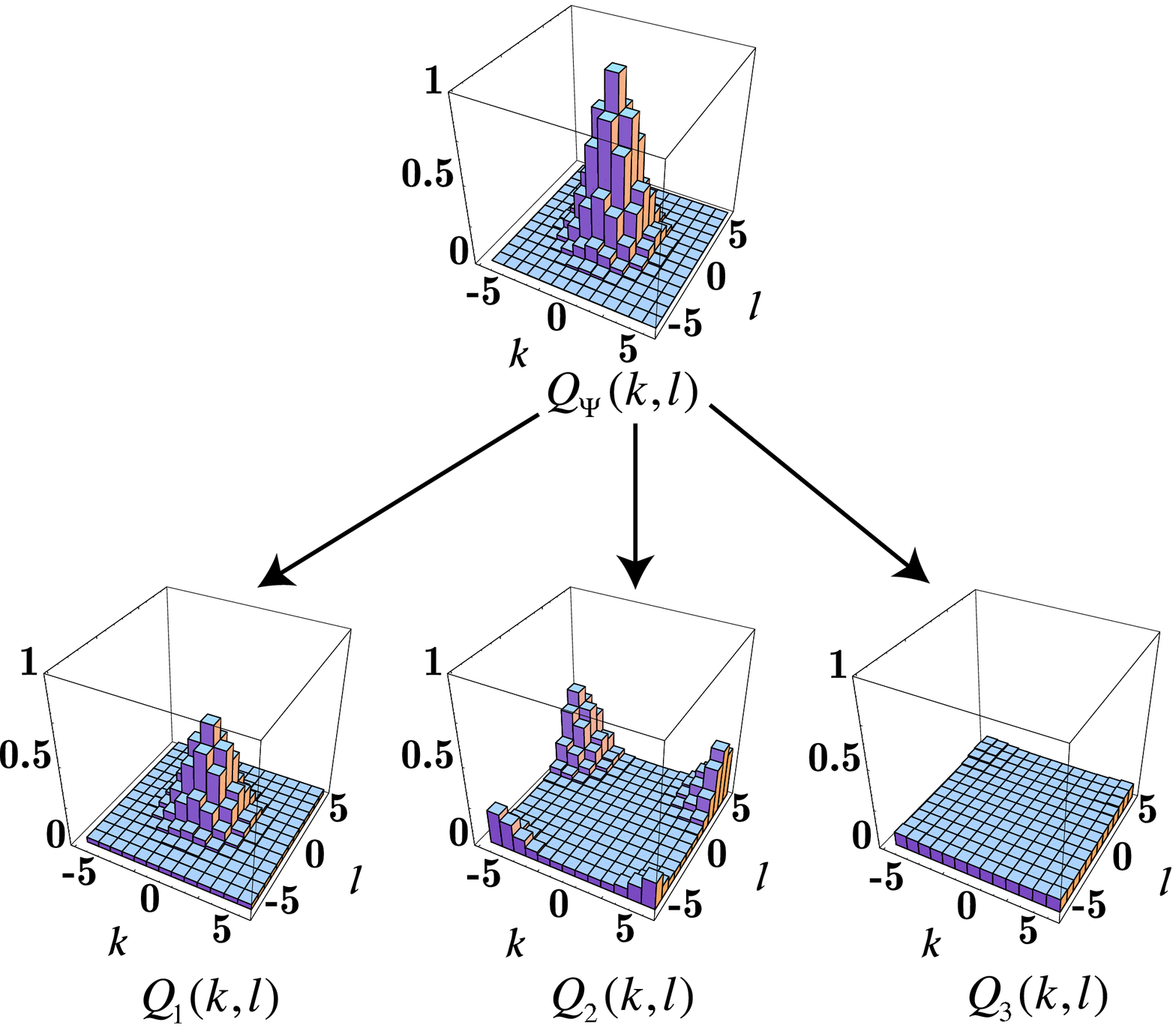}}
\medskip
\caption{ Husimi functions of the input state of the data qudit
and the output qudits. The input data qudit is initially prepared
in the ground state of the Hamiltonian (\ref{4.10}) while the
auxiliary system (ancilla) is initially prepared in the state
$|\Theta\rangle=0.75|\Xi_{00}\rangle-0.64|x_7 \rangle|p_5 \rangle$.
The top graph, labeled  $Q_{\Psi}(k,l)$,
represents the Husimi function of the initial state of the data
qudit.  The three graphs, labeled  $Q_1(k,l)$, $Q_2(k,l)$ and
$Q_3(k,l)$, represent the Husimi functions of reduced states
$\hat{\rho}_1$, $\hat{\rho}_2$ and $\hat{\rho}_3$ of the composite
system that are given by Eqs.~(\ref{5.3})-(\ref{5.5}),
respectively. } \label{fig3}
\end{figure}

In order to illustrate the action of the QID we plot in
Fig.~\ref{fig3} $Q$ functions of an input qudit, that is,
initially prepared in the ground state of Hamiltonian
(\ref{4.10}), as well as the three output qudits. The ruler state
is chosen to be again the ground state of Hamiltonian
(\ref{4.10}). The Husimi functions do correspond to the situation
when a POVM measurement is performed on the density operator
$\hat{\rho}_j$ ($j=1,2,3$) given by Eqs. (\ref{5.3}) --
(\ref{5.5}), respectively.

\end{multicols}
\vspace{-0.2cm}
\noindent\rule{0.5\textwidth}{0.4pt}\rule{0.4pt}{0.6\baselineskip}
\vspace{0.2cm}
  \hfill\noindent\rule[-0.6\baselineskip]%
  {0.4pt}{0.6\baselineskip}\rule{0.5\textwidth}{0.4pt}
\vspace{-0.2cm}
\begin{multicols}{2}

\section{EFFECT OF MEASUREMENT}

It is obvious from expression (\ref{5.4}) that if the von Neumann
measurement using the projector $|\Phi\rangle_2\langle\Phi|$
(i.e., projecting on the ruler state) on qudit 2 is performed then
this measurement results in a reconstruction of the Husimi
function of the original data state affected by the amount of
noise determined by the particular value of $\alpha$. In other
words, this projective measurement will result in the
reconstruction of the Husimi function of the operator
$\hat{\rho}_2^{\rm
(out)}=(1-\alpha^2)\hat{\rho}+\frac{\alpha^2}{N}\hat{\openone}$.
Certainly, the state of the data register is then affected not
only by the action of the QID but also by the effect of the
projective measurement performed on the second qudit.

\begin{figure}[tbp]
\centerline {\epsfig{width=8.0cm,file=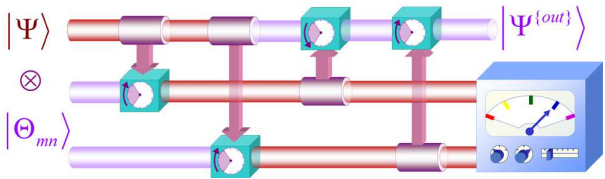}}
\medskip
\caption{Logical network for the quantum information distributor
with a projective measurement performed on the program register.
}
\label{fig4}
\end{figure}

To understand the role of the projective measurement performed on
the program register on the state of the data register at the
output of the QID, let us consider the following. We will study
the action of the quantum-information distributor when the two
ancillary qudits are prepared in a superposition state given by
Eq.~(\ref{5.1}) With this program state the QID acts on the input
data qudit $|\Psi\rangle_1=\sum_k c_k |x_k\rangle$ so that at the
output the three qudits are in the state:
 \be
\label{6.1} |\Omega^{(out)}\rangle_{123} &=&
U_{123}|\Psi\rangle_1\left[\alpha |\Xi_{00}\rangle_{23} +\beta
|x_m\rangle_2|p_n\rangle_3\right]
\\
&=&\alpha |\Psi\rangle_1 |\Xi_{00}\rangle_{23}
+\beta\left[\hat{R}_x(m)\hat{R}_p^\dagger(n)
|\Psi\rangle\right]_2 |\Xi_{nm}\rangle_{31}
\; .
\nonumber
 \ee
Then we will assume that both program qudits are measured
projectively. Qudit 2 is projected in the ruler state
$|\Phi\rangle_2=\sum_k f_k|x_k\rangle_2$ while the qudit 3 is
projected on the transposed ruler state $|\Phi^{\rm
T}\rangle_3:=\sum_k f_k^*|x_k\rangle_3$. Schematically this
situation is depicted in Fig.~\ref{fig4}

The data qudit after the action of the QID and this projective
measurement reads \be \label{6.2}
 &&|\Psi^{(out)}\rangle_1 \simeq
_2\langle\Phi|\, _3\langle\Phi^{\rm T}|
\Omega^{(out)}\rangle_{123}\\
&&= \frac{\alpha}{\sqrt{N}}|\Psi\rangle_1 + \frac{\beta \langle
\Phi|\hat{R}_x(m)\hat{R}_p^\dagger(n)|\Psi\rangle}{\sqrt{N}}
\hat{R}_x^\dagger(m)\hat{R}_p(n) |\Phi\rangle_1\; . \nonumber \ee
This means that by acquiring knowledge of a particular value of
the Husimi functions of the second and the third qudits, the data
qudit ``collapses'' into the state (\ref{6.2}). The disturbance of
the original data state depends on the value of $\alpha$, the
particular point $(m,n)$ at which the Husimi functions of the
program qudits are measured and the specific choice of the ruler
state.

\section{CONCLUSION}

Here we have shown how a simple quantum device, the
quantum-information distributor, can both distribute and process
quantum-information.  This device was discussed in Ref.\
\cite{Braunstein2001}, and it was shown there that the flow of
information was controlled by a program state. In this paper, we
have considered a much wider class of program states. Besides
moving the quantum information between outputs, they allow us to,
in addition, apply shift operators to the input data.  This, in
turn, makes it possible to use the QID to measure the discrete $Q$
function of the input data, which is equivalent to realizing a
class of POVM operators.  Another possibility, is to split the
input into two parts, find the $Q$ function of one part and retain
the other part. There is a trade-off involved: the more
information that is retained, the more smeared is the $Q$
function, and the better the $Q$ function, the more distorted is
the information in the retained qudit.  Thus, the QID provides us
with a very flexible programmable quantum-information processing
device, which has a number of useful applications.

\acknowledgements This work was supported in part by the European
Union  project QGATES and by the grant agency VEGA of the Slovak
Academy of Sciences and by NSF grant No. PHY-0139692. V.B. would
like to thank the Science Foundation Ireland for support.

\end{multicols}
\end{document}